# Reflected wavefront manipulation by acoustic metasurface with anisotropic local resonant units

Pai Peng[(a)], Meiyu Liu and Pan Li

*School of Mathematics and Physics, China University of Geosciences- Wuhan, China*

**Abstract** –Metasurfaces with planar profile and wave front shaping capabilities would be ideally suitable to improve the performance of acoustic wave-based applications. It is significant that the general Snell's law provides a new approach to engineer the phase profile of reflected acoustic wave. Here, we present a new type of acoustic metasurface based on three-component composites which contain a steel ball coated with a thin elliptical layer of silicone rubber and embedded in epoxy. The acoustic metasurface with changeable radii of steel balls can realize the anomalous reflections, a planar acoustic lens through designed gradient phase profile. The acoustic metasurface contained with elaborately selected three-component composites, as a skin cloak, successfully achieve the acoustical invisibility.

Recent years have witnessed a great development of acoustic metamaterials with various structures, which exhibit novel acoustic properties that can never be realized by natural materials in controlling sound waves, such as negative mass density, negative modulus and double-negative parameters[1-6]. However, these metamaterials have large thickness, leading to bulky size which goes against miniaturization and integration of acoustic devices. A new type of metamaterial, namely, metasurfaces with compact planar subwavelength structure have advantages over many forms of novel wave front manipulation and are attracted significant attention by physics and engineering communities. Not only do acoustic metasurfaces exhibit the ability of governing the acoustic wave by the general Snell's law, but also they provide a new method to accomplish the integration of acoustic devices. The general Snell's law is presented to rich the anomalous reflection and refraction phenomenon with phase discontinuities. The specific metasurfaces display the novel acoustic wave front control phenomena, such as anomalous reflection and refraction, propagated wave converting to surface wave, lens, have been well carried out in the theoretically and experimentally, applying the structures that contain the Helmholtz resonant array [7-9] ,tapered labyrinthine structure [10-12], coiling-up slit structure [13-16] and membrane based structure [17]. Li et al. [7] utilized a metascreen composed of elements with four Helmholtz resonators (HRs) in series and a straight pipe to realize the anomalous refraction even if the incident wave is obliquely propagated with large angle. Xie et al.[11] applied the tapered labyrinthine metamaterial that have high impedance matching to firstly realize conversion from propagating wave to surface mode, extraordinary beam-steering and apparent negative refraction through higher-order diffraction. Li et al. [15] designed a coiling-up metasurface with discrete phase shift covering the full 2π span to realize the arbitrary modulation of phase profiles. Chen [17] also designed an acoustic composite structure of cavity and membrane to manipulate the phase profile in a controllable manner at low frequency. Meanwhile, acoustic metasurfaces can be utilized in skin cloak [18-21]. Zhai et.al [20] designed 2D acoustic ultrathin skin cloaks constructed by a cavity coupled with a membrane, which entirely compensate the wave front discrepancy generated by the scattering of the hidden object. Xu et al. [21] proposed the acoustic metasurface made of graded spiral units. An acoustical layer consisting of 80 subwavelength-sized unit cells is selected to provide precise local phase modulation, making the object invisibility possible. However, these metasurfaces with different inner structures can only propagate the sound wave in the fluid matrix. The three-component composites provide a possibility of changing the direction of controllable elastic wave in the solid matrix. Now we start to introduce the three-component composites in the fluid matrix. And they are also capable of controlling reflected acoustic waves arbitrarily and open up a new avenue for acoustic wave front engineering and manipulations.

In this paper we utilize another acoustic resonant structure such as three-components to realize a low frequency acoustic metasurface, which can realize arbitrarily steering the reflected wave. The structure contains the three accessible materials, which demonstrate remarkable phase change and novel acoustic property. Anomalous reflection, flat acoustic aberration-free lens acoustic cloaking are exhibited by imposing suitable phase delay profiles of the metasufaces. The designed metasurfaces extend new degrees of freedom and

[(a)]E-mail: paipeng@cug.edu.cn
China University of Geoscience - Wuhan China.

paves the way for thin planar reflective structures for acoustic wave front manipulation.

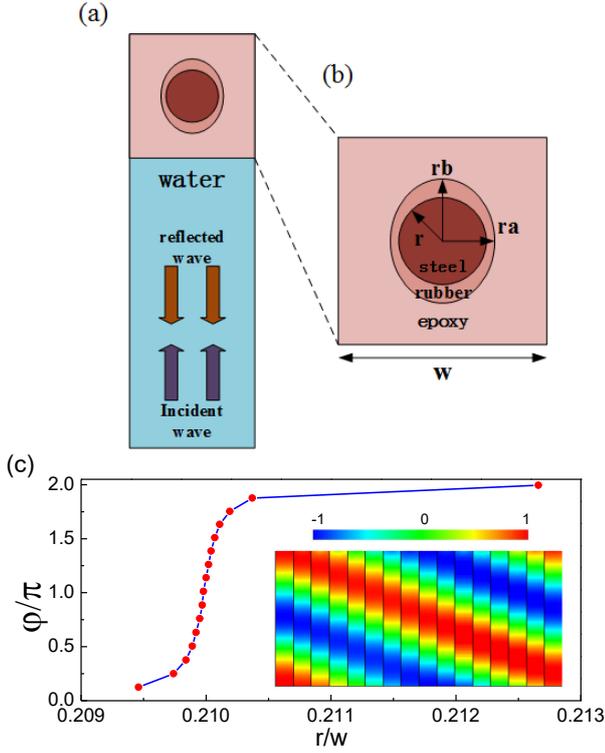

Fig. 1: (a) Schematic of the proposed three-component subunit, which comprises a steel ball coated with a thin elliptical layer of silicone rubber and embedded in epoxy. The dark purple and orange arrows refer to the propagation direction of incident wave and reflected wave, respectively. (b) The radius of steel ball r is an adjustable parameter and the long and short axes of ellipse are constant. The three-component structure has a width w=1m and the length is same. Prescribed velocity conditions are imposed to the three boundaries to keep each other independent. (c) The phase of reflected wave, as a function of the radii of steel ball, with the incident frequent $f_0 = 169.3 Hz$ is plotted .The pressure strips of the reflected wave are shown for 16 three-component subunits.

Figure 1(a) shows the schematic of unit for the planar acoustic metasurface. The basic resonant unit is composed of a steel ball coated with a thin elliptical layer of silicone rubber and embedded in epoxy. The parameters of materials used are: $\rho_e = 1180\,Kg/m^3$ , $\lambda_e = 4.4 \times 10^9\,N/m^2$ and $\mu_e = 1.6 \times 10^9\,N/m^2$ for epoxy; $\rho_r = 980\,Kg/m^3$ , $\lambda_r = 1.96 \times 10^9\,N/m^2$ and $\mu_r = 5.5 \times 10^5\,N/m^2$ for rubber; $\rho_s = 7900\,Kg/m^3$ , $\lambda_s = 1 \times 10^{11}\,N/m^2$ and $\mu_s = 8.1 \times 10^{10}\,N/m^2$ for steel, respectively, where $\rho$ is mass density, $\lambda$ and $\mu$ are Lame constants. To present the phase delay profile with a relatively high resolution, 16 types of three-component composites with trihedral rigidity were designed and optimized. So as to forming a planar acoustic metasurface and for the simplicity of design, the radii of steel balls are tailored to yield the desired phase shifts, while the values of two axis of ellipse are fixed in the calculation. The two axis of ellipse are 0.25m and 0.30m, respectively. The three-component structure has a width w=1m and the length is same. The phase of the reflected waves as a function of the radii of steel ball is plotted in Figure1. (c). It is significant that the sixteen designed units are able to realize $2\pi$ phase spans with an interval of $\pi/8$ between the adjacent subunits, denoted by red dots. To further illustrate the properties of 16 three-component units, we present reflected acoustic pressure at 169.3Hz as plotted in Figure 1c. The coated elliptical rubber acting as spring to connect the stiff steel ball and epoxy have a small longitudinal wave speed. Since the rubber have a comparable size with the incident wavelength, making the coated elliptical rubber and steel ball to vibrate as a whole along the direction of wave propagation [23-24]. Compared to the sound waves propagating in the complete water medium with the same length, there exists substantial phase delay as the ones travel through the three-component composites, which shows potential to manipulate the phase of reflected acoustic wave in space. When we select the appropriate reflected phase with specific steel ball radii, each unit is individually simulated to apply both periodic boundary condition and the trihedral rigidness to decrease the boundary effect and keep independent with the adjacent units. In addition, the thin elliptical rubber have a big advantage over stability to keep the sound waves propagating in specific direction, owing that the three-components with the round one impinged by the sound waves are likely to form forced vibrations in the different directions. From the above statement, setting the three-component composites (a kind of resonant structure) as building blocks shows a promise for wavefront-shaping metasurfaces.

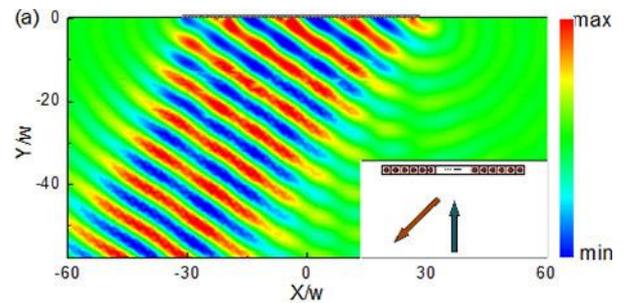

Fig. 2: (a) Pressure field pattern of the reflected acoustic wave for gradient phase profile is $d\varphi(x)/dx = \pi/8w$, when the acoustic waves impinge the metasurface from the bottom to the top.

To design an acoustic metasurface with nearly arbitrary wave front modulation and planar profile, some particular units should be selected based on a series of three-component composites under the guidance of the general Snell's law. Compared with the classical Snell's law which depends on the continuous accumulated phase, the general Snell's law with the abrupt phase can similarly govern the direction of reflected acoustic waves as an incident plane wave with angle $\theta_i$ impinges the acoustic metasurface. The formula is as follows [22]:

$$\sin(\theta_r) - \sin(\theta_i) = \frac{d\varphi(x)}{k_0 n_i dx} \quad (1)$$

Where $\theta_r$, $\theta_i$ are the reflected and incident angles, respectively; $\varphi(x)$, $dx$ are the phase discontinuities and the distant between the crossing points along the x direction, respectively; $k_0 = 2\pi/\lambda$ is the wave vector in water; $n_i$ denotes the refractive indices of water. Equation (1) shows the direction of reflected wave can be controlled freely through a suitable gradient phase profile.

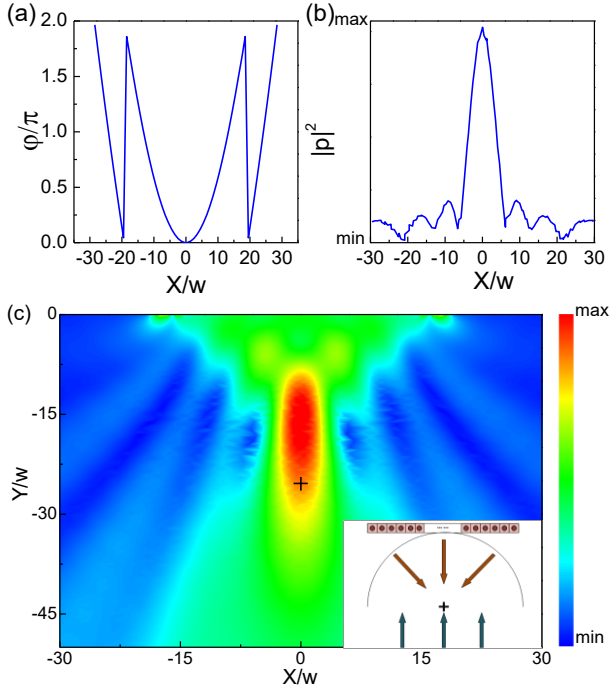

Fig. 3: (a) A desire hyperboloidal phase profile along x direction of reflected wave for the designed lens. (b) The transverse distribution of acoustic pressure intensity profile at y=20.5m. (c) The pressure field distribution of reflected wave for the designed metasurface with $f = 3\lambda$, as the incident acoustic waves propagate from the bottom.

To support the validity of the three-component composites about anomalous reflection, numerical simulations of the constructed metasurfaces were introduced imposing the specific phase gradients. When an acoustic wave from the bottom is normally incident toward the designed metasurfaces, the reflected pressure fields show the acoustic wave obliquely. In Figure 2(a), it is worth pointing out that the phase gradient is set as $d\varphi(x)/dx = \pi/8w$ and the angle of reflected wave is calculated as 42.5° utilizing the general Snell's law with the normal incident. From the reflected pressure distribution it is noted that the simulated results agree well with the theoretical ones. It is necessary to mention that the local resonance of three-component composites have a contribution to the anomalous reflection and provide a feasible thought to design the acoustic metasurface.

Then, to further verify the three-component composites to satisfy the capacity of wave front modulation, we present a flat lens by well-designed acoustic metasurface. It is no doubt that controlling the phase profile is circle by delaying the acoustic phase employing the elaborate designed three-component composites to realize the acoustic focusing. For a given focal length f, the phase shift $\varphi(x)$ imposed at every point must satisfy the following equation:

$$\varphi(x) = k_0\sqrt{x^2 + f^2} - f \quad (2)$$

Applying the aforementioned condition, when the focal length is set as $3\lambda$, the structure of three-component composites along any x position is certain. Fig. 3(a) illustrates the continuous phase shifts along the x axis of reflected wave. The spatial reflected pressure intensity distribution is shown in Fig. 3(c). To better exhibit the performance of the acoustic lens, the transverse cross-section pressure intensity distribution at y=20.5 m is plotted in Fig. 3(b).

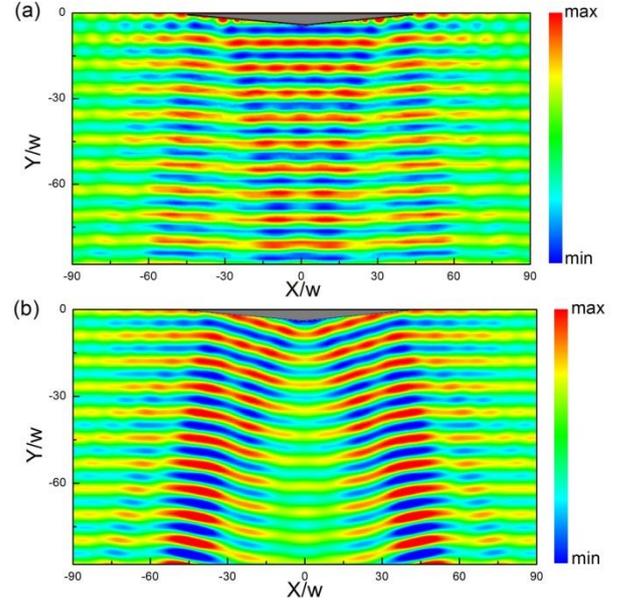

Fig. 4: (a) Pressure field of the object covered by the selected three-component composites exposed to the normally acoustic wave. (b) Pressure field of the uncloak object exposed to the normally acoustic wave.

Apart from the extraordinary ability of manipulating the acoustic wave, acoustic metasurfaces also have an advantage over shielding an object from incident wave. Compared with the bulky cloaks based on coordination transformation, cloaks with metasurface that have a smaller size can simultaneously control the wave amplitude and the wave phase. The designed three-component composites provide precise phase modulation and hence remain the scattered acoustic field invariability in such a way to achieve invisibility. We choose the width and height of the triangular object as $4\lambda$ and $0.5\lambda$, respectively, being covered the inclined plane with the three-component composites. Three-component composites along the surface of triangular object are elaborately selected according to the height of triangular object at the corresponding position. To verify the performance of the designed skin cloak, numerical simulations are carried out for the cases when acoustic wave impinges normally from the bottom toward the uncloaked object and the object covered by the metasurface. In fig. 4(a), it is obvious that the total acoustic pressure field exhibits the well performance for invisibility no matter what the object is, owing to the efficient phase modulation. However, the uncloaked object induces a strong perturbation to the scattering acoustic pressure field, in fig. 4(b). Nearly perfect cloaking can be observed by comparing the scattering field distributions without and with metasurface based on the three-component composites.

In conclusion, we present a new sort of acoustic metasurfaces comprised of three-component composites with changeable radii of steel balls or adjustable deflective angle of ellipses, which manipulates the wave front arbitrary based on the general Snell's law. The acoustic metasurface with changeable radii of steel balls can realize the anomalous reflections through different gradient phase profile. Furthermore, a planar acoustic lens is presented applying the hyperbolic phase profile. The acoustic metasurface constructed by three-component composites have a great potential in cloaking, realizing the triangular object invisibility from the scattering acoustic pressure field. Our designed metasurfaces provide a new approach for wave front engineering and manipulation. Three-component composites have a great potential in manipulating the elastic wave in the solid matrix.

\*\*\*

This work was supported by the National Natural Science Foundation of China (Grant No: 11604307)